\documentstyle[psfig,eps]{mn}

\newif\ifAMStwofonts

\def\simgt{\hbox{\rlap{\raise 0.425ex\hbox{$>$}}\lower 0.65ex\hbox{$\sim$}}}
\def\simlt{\hbox{\rlap{\raise 0.425ex\hbox{$<$}}\lower 0.65ex\hbox{$\sim$}}}
\def\h1{$h^{-1}$}

\ifoldfss
  \ifCUPmtlplainloaded \else
    \NewTextAlphabet{textbfit} {cmbxti10} {}
    \NewTextAlphabet{textbfss} {cmssbx10} {}
    \NewMathAlphabet{mathbfit} {cmbxti10} {} 
    \NewMathAlphabet{mathbfss} {cmssbx10} {} 
  \fi
  \ifAMStwofonts
    \ifCUPmtlplainloaded \else
      \NewSymbolFont{upmath} {eurm10}
      \NewSymbolFont{AMSa} {msam10}
      \NewMathSymbol{\upi}     {0}{upmath}{19}
      \NewMathSymbol{\umu}     {0}{upmath}{16}
      \NewMathSymbol{\upartial}{0}{upmath}{40}
      \NewMathSymbol{\leqslant}{3}{AMSa}{36}
      \NewMathSymbol{\geqslant}{3}{AMSa}{3E}

      \let\leq=\leqslant 
       
    \fi
  \fi
\fi 

\ifnfssone
  \newmathalphabet{\mathit}
  \addtoversion{normal}{\mathit}{cmr}{m}{it}
  \addtoversion{bold}{\mathit}{cmr}{bx}{it}
  \newmathalphabet{\mathbfit} 
  \addtoversion{normal}{\mathbfit}{cmr}{bx}{it}
  \addtoversion{bold}{\mathbfit}{cmr}{bx}{it}
  \newmathalphabet{\mathbfss} 
  \addtoversion{normal}{\mathbfss}{cmss}{bx}{n}
  \addtoversion{bold}{\mathbfss}{cmss}{bx}{n}
  \ifAMStwofonts
    \ifCUPmtlplainloaded \else
      \UseAMStwoboldmath
      \makeatletter
      \new@mathgroup\upmath@group
      \define@mathgroup\mv@normal\upmath@group{eur}{m}{n}
      \define@mathgroup\mv@bold\upmath@group{eur}{b}{n}
      \edef\UPM{\hexnumber\upmath@group}
      \new@mathgroup\amsa@group
      \define@mathgroup\mv@normal\amsa@group{msa}{m}{n}
      \define@mathgroup\mv@bold\amsa@group{msa}{m}{n}
      \edef\AMSa{\hexnumber\amsa@group}
      \makeatother
      \mathchardef\upi="0\UPM19
      \mathchardef\umu="0\UPM16
      \mathchardef\upartial="0\UPM40
      \mathchardef\leqslant="3\AMSa36
      \mathchardef\geqslant="3\AMSa3E

      \let\leq=\leqslant 

    \fi
  \fi
\fi 
\ifnfsstwo
  \DeclareMathAlphabet{\mathbfit}{OT1}{cmr}{bx}{it}
  \SetMathAlphabet\mathbfit{bold}{OT1}{cmr}{bx}{it}
  \DeclareMathAlphabet{\mathbfss}{OT1}{cmss}{bx}{n}
  \SetMathAlphabet\mathbfss{bold}{OT1}{cmss}{bx}{n}
  \ifAMStwofonts
    \ifCUPmtlplainloaded \else
      \DeclareSymbolFont{UPM}{U}{eur}{m}{n}
      \SetSymbolFont{UPM}{bold}{U}{eur}{b}{n}
      \DeclareSymbolFont{AMSa}{U}{msa}{m}{n}
      \DeclareMathSymbol{\upi}{0}{UPM}{"19}
      \DeclareMathSymbol{\umu}{0}{UPM}{"16}
      \DeclareMathSymbol{\upartial}{0}{UPM}{"40}
      \DeclareMathSymbol{\leqslant}{3}{AMSa}{"36}
      \DeclareMathSymbol{\geqslant}{3}{AMSa}{"3E}

      \let\leq=\leqslant 

    \fi
  \fi
\fi 

\ifCUPmtlplainloaded \else
  \ifAMStwofonts \else 
    \def\upi{\pi}
    \def\umu{\mu}
    \def\upartial{\partial}
  \fi
\fi

\title[Cosmological Obscuration by Galactic Dust]
{Cosmological Obscuration by Galactic Dust: Effects of Dust Evolution} 
\author[F. J. Masci \& R. L. Webster]
{Frank J. Masci$^1$\thanks{Email: {\bf \tt fmasci@ipac.caltech.edu}}
and Rachel L. Webster$^2$\thanks{Email: 
{\bf \tt rwebster@physics.unimelb.edu.au}}
\\$^1$Infrared Processing and Analysis Center, M/S 100-22, 770 South
Wilson Avenue, 
\\~California Institute of Technology, Pasadena, CA 91125 
\\$^2$School of Physics, University of Melbourne,
Parkville, Victoria 3052, Australia}
\date{Zeroth Draft} 
\pubyear{1998}

\begin{document}

\maketitle

\newcommand{\fmmm}[1]{\mbox{$#1$}}
\newcommand{\scnd}{\mbox{\fmmm{''}\hskip-0.3em .}}
\newcommand{\scnp}{\mbox{\fmmm{''}}}

\begin{abstract}
We explore the effects of dust in cosmologically distributed
intervening galaxies on the high redshift universe using a generalised
model where dust content evolves with cosmic time. 
The absorbing galaxies are modelled as exponential disks
which form coevally, maintain a constant space density and evolve in
dust content at a rate that is uniform throughout. 
We find that
the inclusion of moderate to moderately weak amounts of evolution 
consistent with other studies
can reduce the mean observed $B$-band optical depth to redshifts $z\simgt1$
by at least 60\% relative to non-evolving models.
Our predictions imply that intervening galactic dust is unlikely to bias 
the optical counts of quasars at high redshifts 
and their evolution in space density derived therefrom. 
\end{abstract}


\section{Introduction}

The recent discovery of large 
numbers of quasars at radio and X-ray frequencies with 
very red optical--to--near-infrared continua
suggests that existing optical surveys may be severely incomplete
(eg. Webster et al. 1995 and references therein). 
Webster et al. (1995) and Masci (1998) have argued that the anomalous 
colours are due to extinction by dust,
although the location of the dust remains 
a highly controversial issue.
Intervening dusty galaxies which happen to lie along the line-of-sight of
otherwise normal blue quasars are expected to redden the observed
optical continuum, or if the optical depth is high enough, to
remove quasars from an optical flux-limited sample (eg. Wright 1990).  
As suggested by existing obervational and theoretical 
studies of cosmic chemical evolution however (Pei \& Fall 1995 and
references therein), one expects a reduction in the
amount of dust to high redshift. Consequently, one then also expects
that the probability of a
background object being either reddened or obscured to be reduced. 

The effects of foreground dust on observations 
of objects at cosmological distances
has been discussed by Ostriker \& Heisler (1984); Heisler \& Ostriker
(1988); Fall \& Pei (1989, 1993); Wright (1986, 1990) and 
Masci \& Webster (1995).
Using models of dusty
galactic disks,
these studies show that the line-of-sight
to a high redshift quasar has a high probability of being
intercepted
by a galactic disk, particularly if the dust distribution is larger than the
optical radius of the galaxy. 
Based on the dust properties of local galaxies, 
it is estimated that up to 80\% of bright quasars 
to $z\sim3$ may be obscured by dusty intervening systems.
The principle issue in these calculations was that 
realistic
dust distributions in galaxies which are `soft' around the
edges, will cause many quasars to appear 
reddened without
removing
them from a flux-limited sample.

None of the above studies however considered the effects of evolution
in dust content.
Cosmic evolution in dust is indirectly suggested by numerous 
claims of reduced chemical enrichment at $z\simgt2$.
Evidence is provided by observations of trace metals
and their relative abundances in QSO
absorption-line systems to $z\sim 3$ (Meyer \& Roth 1990;
Savaglio, D'Odorico \& M\"{o}ller 1994; Pettini et al. 1994; Wolfe
et al. 1994;
Pettini et al. 1997; Songaila 1997),
which are thought to arise
from intervening clouds or the haloes and 
disks of galaxies. These studies indicate
mean metallicities $\simeq 10\%$ and $\simlt1\%$ 
solar at $z\sim2$ and $z\sim3$ respectively, and dust-to-gas ratios
$\simlt8\%$ of the galactic value at $z\sim2$.
These estimates are consistent with simple global evolution models of 
star formation and gas consumption rates
in the universe 
(Pei \& Fall 1995).
If the observed metallicities in QSO absorption systems are common, 
then their interpretation
as galactic disks implies that substantial evolution
has taken place since $z\sim 3$.
If the quantity of dust on cosmic scales also follows such a trend, 
then one may expect
the effects of
obscuration to high redshift to be reduced relative to
non-evolving predictions.

In this paper, we continue to model the effects of
intervening galactic dust
on the background universe at optical wavelengths using a more 
generalised model where the dust content evolves. 
We explore the effects of our predictions on quasar number counts in 
the optical and their implication for quasar evolution. 
 
This paper is organised as follows:
The next section briefly describes the generalised model and assumptions. 
Section~\ref{mprev}
describes the model parameters and their values assumed in our calculations. 
Model results are presented and analysed in Section~\ref{rests}. 
Implications on quasar statistics and evolution
are discussed in Section~\ref{QSOev}. 
Other implications are discussed in Section~\ref{evmodd} and 
all results are summarised in Section~\ref{concfour}.
Unless otherwise stated, all calculations assume a Friedmann cosmology with
$q_{0}=0.5$, and Hubble parameter $h_{50}=1$ 
where $H_{0}=50h_{50}\, \rm km\,s^{-1}\,Mpc^{-1}$.

\section{The Evolutionary Dust Model}
\label{modeldes}
 
We calculate the probability distribution in total dust optical depth from model
galaxies along any random line-of-sight as a function of redshift by
following the method presented in Masci \& Webster (1995). 
This was based on a method introduced by 
Wright (1986) which did not 
include any effects of evolution with redshift. Here we generalise
this model by considering the possibility of evolution in the dust
properties of galaxies.
In the discussion below and unless otherwise indicated by a
subscript, we define $\tau$ to be 
the total optical depth 
encountered by emitted photons and 
measured in an {\it observer's $B$ bandpass} (effectively at $\lambda=4400$\AA).
 
We assume the following properties for individual
absorbing galaxies. Following previous studies (eg. Wright 1986, Heisler \&
Ostriker 1988), we model 
galaxies as randomly tilted exponential disks,
where the
optical depth through a face-on disk decreases exponentially
with distance $r$ from the center:
\begin{equation}
\tau(r,z)\,=\,\tau_{0}(z)\,e^{-r/r_{0}}.
\label{expr}
\end{equation}
$r_{0}$ is a characteristic radius and $\tau_{0}(z)$,
the value of $\tau$ through the center of the galaxy $(r=0)$.
The redshift dependence of $\tau_{0}$ is due to the increase in 
absorber rest frame frequency with redshift. 
 
Since we wish to model the observed $B$-band optical depth to $z\simlt6$, 
we require an extinction law $\xi(\lambda)\equiv\tau_{\lambda}/\tau_{B}$ that
extends to wavelengths of $\sim630$\AA~. 
We use the analytical fit for $\xi(\lambda)$ as derived by Pei (1992) for
diffuse galactic dust in the range $500{\rm\AA}\simlt\lambda\simlt25\mu$m. 
The optical depth in an observer's frame 
through an absorber at redshift $z$ ($\tau_{0}(z)$ in equation~\ref{expr}) 
can be written:
\begin{equation}
\tau_{0}(z)\,=\,\tau_{B}\,\xi\left(\frac{\lambda_{B}}{1+z}\right),
\label{tz}
\end{equation}
where $\tau_{B}$ is 
the {\it rest frame} $B$-band optical depth through the
center of an individual galactic absorber.

\subsection{Evolution} 
\label{ev}
 
Equation (\ref{tz}) must be modified if the dust content in each
galaxy is assumed to evolve with cosmic time. 
The optical depth seen through the
center of a single absorber at some redshift, $\tau_{0}(z)$, 
will depend
on the quantity of
dust formed from past stellar processes.
For simplicity, we assume all galaxies form simultaneously, maintain a
constant space density,
and increase in dust content at a rate that is 
uniform throughout.
We also assume no evolution in the dust law $\xi(\lambda)$ with redshift.
Even though a lower mean metallicity at high redshift 
may suggest a different wavelength dependence for 
the dust law, there is no evidence
from local observations of the diffuse ISM to support this view 
(eg. Whittet 1992). 
 
We parameterise evolution in dust content by following
simulations of the formation of heavy metals in the
cold dark matter scenario of galaxy formation by  
Blain \& Longair (1993a, 1993b).
These authors assume that galaxies form
by the coalescence of gaseous protoclouds through
hierarchical clustering as prescribed by Press \& Schechter (1974).
A fixed fraction of the mass involved in each merger event is converted
into stars, leading to the formation
of heavy metals and dust.
It was assumed that 
the energy liberated
through stellar radiation was absorbed
by dust and re-radiated into the far-infrared. 
They found that such radiation can contribute substantially
to the far-infrared background intensity from which they use to constrain 
a model
for the formation of heavy metals
as a function of cosmic time.
Their models show that the comoving density of heavy metals 
created by some redshift $z$, given that star formation
commenced at some epoch $z_{SF}$ follows the form 
\begin{equation}
\Omega_{m}(z)\,\propto\,\ln\left({1+z_{SF}\over1+z}\right),
{\rm\hspace{6mm} where}\,z<z_{SF}.
\label{omegaZ}
\end{equation}

We assume that a fixed fraction of heavy metals condense
into dust grains so that
the comoving density in dust, $\Omega_{d}(z)$,
follows a similar dependence as equation (\ref{omegaZ}).
The density in dust relative to the present closure density
in $n_{0}$ exponential
disks
per unit comoving volume is given by 
\begin{equation}
\Omega_{d}\,=\,\frac{n_{0}M_{d}}{\rho_{c}},
\label{omegaD}
\end{equation}
where $\rho_{c}=3H_{0}^{2}/8\pi G$ and 
$M_{d}$ is the dust mass in a single exponential disk. 
This mass can be estimated using Eq.7-24 from Spitzer (1978)
where the total density in dust, $\rho_{d}$,
is related to the extinction $A_{V}$ along a path length $L$ in kpc
by 
\begin{equation}
\langle\rho_{d}\rangle\,=\,1.3\times10^{-27}\rho_{g}
\left(\frac{\epsilon_{o} + 2}{\epsilon_{o} - 1}\right)\,(A_{V}/L).
\label{rhod}
\end{equation}
$\rho_{g}$ and $\epsilon_{o}$ are the density and
dielectric constant of a typical dust grain respectively 
and the numerical factor has
dimensions of
${\rm gm\,cm}^{-2}$ - see Spitzer (1978).
Using the exponential profile (equation \ref{expr}) where
$\tau(r)\propto A_{V}(r)$
and integrating along cylinders, 
the
dust mass in a single exponential 
disk can be found in terms of the model parameters
$\tau_{B}$ and $r_{0}$.
We find that the comoving density in dust at some redshift scales as
\begin{equation}
\Omega_{d}(z)\,\propto\,\tau_{B}(z)\,n_{0}\,r_{0}^{2},
\label{omD2}
\end{equation}
where $\tau_{B}(z)$ is the central $B$-band
optical depth and $r_{0}$ the dust scale radius 
of each disk.
Thus, the central optical depth, $\tau_{B}(z)$, in any
model absorber at some redshift
is directly proportional to the mass density in dust or 
heavy metals
as specified by equation (\ref{omegaZ}):
\begin{equation}
\tau_{B}(z)\,\propto\,\ln\left({1+z_{SF}\over1+z}\right).
\label{tbz}
\end{equation}
 
\begin{figure}
\vspace{-1in}
\plotonesmall{1.1}{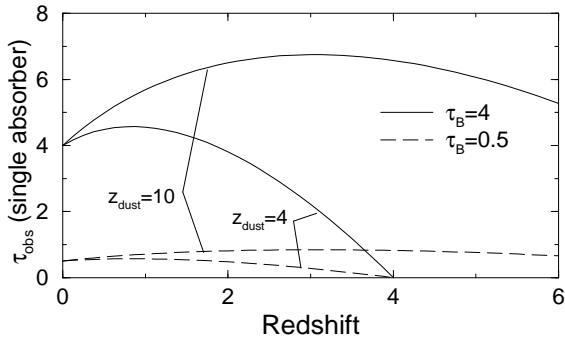}
\vspace{-1.7in}
\caption[``Observed'' optical depth through a single model absorber]{
Optical depth in an observer's $B$-bandpass as a function 
of redshift through a single model absorber defined by equation (\ref{tboz}).
$\tau_{B}$ is the rest frame central $B$-band optical depth and 
$z_{dust}$ the dust formation epoch.
}
\label{singleabs}
\end{figure}

The redshift dependence of optical depth observed in the fixed
$B$-bandpass due to a {\it single} absorber now involves two factors:
first, the extinction properties of the dust as defined by equation (\ref{tz}) 
and second, its evolution specified by equation (\ref{tbz}).
The star formation epoch $z_{SF}$ can also be interpreted as 
the redshift at which dust forms. From here on, we 
therefore refer to this parameter 
as $z_{dust}$ - a hypothesised ``dust
formation epoch''.
By convolving equations (\ref{tz}) and (\ref{tbz}), and requiring
that locally: $\tau_{0}(z=0)\,=\,\tau_{B}$,
the {\it observed} optical depth through a {\it single} absorber at some redshift
$z<z_{dust}$ now takes the form: 
\begin{equation}
\tau_{0}(z)\,=\,\tau_{B}\,\xi\left(\frac{\lambda_{B}}{1+z}\right)\left[1 - 
{\ln(1+z)\over\ln(1+z_{dust})}\right].
\label{tboz}
\end{equation}
Figure~\ref{singleabs} illustrates 
the combined effects of evolution and increase in 
observed frame $B$-band extinction with redshift defined by equation (\ref{tboz}).
The extinction
initially increases with $z$ due to a decrease in corresponding 
rest
frame wavelength. Depending on the value for $z_{dust}$, it then
decreases due to evolution in dust content. This latter effect dominates
towards 
$z_{dust}$.
 
The characteristic galactic dust radius $r_{0}$ defined in equation (\ref{expr}) 
is also given a redshift dependence in the sense that galaxies
had smaller dust-haloes at earlier epochs.
The following evolutionary form is adopted:
\begin{equation}
r_{0}(z)\,=\,r_{0}\,(1+z)^{\delta},{\rm\hspace{7mm}}\,\delta < 0,
\label{rozev}
\end{equation}
where $\delta$ gives the rate of evolution and $r_{0}$ is 
now a `local' scale
radius. 
Evolution in radial dust extent is suggested by
dynamical models of
star formation in an initially formed protogalaxy
(Edmunds 1990 and references therein).
These studies show that the star formation rate and
hence metallicity in disk galaxies has a radial dependence
that decreases outwards at all
times. 
It is thus quite plausible that galaxies have an evolving effective
`dust radius' which follows chemical enrichment from stellar
processes.
 
Our parameterisation for evolution in galactic dust (equations \ref{tbz} 
and \ref{rozev}) 
is qualitatively similar to the
`accretion models' for chemical evolution of Wang (1991),
where the effects of grain destruction by supernovae and
grain formation in molecular clouds is taken into account.
The above model is also consistent with empirical age-metallicity
relationships inferred from spectral observations
in the Galaxy (Wheeler, Snedin \& Truran 1989), and models
of chemical evolution on a cosmic scale implied 
by absorption-line observations of quasars 
(Lanzetta et al. 1995; Pei \& Fall 1995). 

\section{Model Parameters and Assumptions}
\label{mprev}
 
\subsection{Model Parameters}
\label{param} 

Our model depends on
four independent parameters which describe the 
characteristics and evolutionary properties of intervening galaxies. 
The parameters defined `locally' are: the comoving number density 
of galaxies $n_{0}$, the characteristic dust radius $r_{0}$, and
dust opacity $\tau_{B}$ at the center of an individual absorber. 
The evolution in $\tau_{B}$ and $r_{0}$ is defined by equations (\ref{tbz}) 
and (\ref{rozev}) respectively. Parameters defining their evolution are
$\delta$ for $r_{0}$, and the 
`dust formation epoch' $z_{dust}$ for $\tau_{B}$.  
Both $n_0$ and $r_0$ have been conveniently combined into the 
parameter $\tau_{g}$ where
\begin{equation}
\tau_{g}\,=\,n_{0}\,\pi r_{0}^{2}\,{c \over H_{0}},
\label{tg2}
\end{equation}
with ${c \over H_{0}}$ being the Hubble length.
This parameter is proportional to the number 
of galaxies and mean optical depth introduced 
along the line-of-sight (see Section~\ref{rests}).
It also represents a
`local' covering factor in dusty galactic disks - the fraction of sky
at the observer covered in absorbers.

In all calculations, we assume a fixed value for $n_{0}$.
From equation (\ref{tg2}), any evolution in the comoving
number density $n_{0}$ is included in the evolution parameter $\delta$
for $r_{0}$ (equation \ref{rozev}).
Thus in general, $\delta$ represents 
an effective evolution parameter for both $r_{0}$ and $n_{0}$. 
Our model is therefore specified
by four parameters: 
$\tau_{g}$, $\tau_{B}$, $\delta$ and $z_{dust}$.
 
\subsection{Assumed Parameter Values}
\label{resan}

Our calculations assume a 
combination of values for the parameters ($\tau_{g}$, $\tau_{B}$) 
and ($\delta$, $z_{dust}$) that bracket the range consistent 
with existing observations. 
The values ($\tau_{g}$, $\tau_{B}$) are chosen
from previous studies of dust distributions and extinction in nearby spirals.
From the studies of
Giovanelli et al. (1994) and Disney \& Phillipps (1995) 
(see also references therein) we assume the range in central optical
depths: $0.5\simlt\tau_{B}\simlt4$, while dust scale radii 
of $5\simlt (r_{0}/{\rm kpc})\simlt30$ are assumed from
Zaritsky (1994) and Peletier et al. (1995). 
For a nominal comoving galactic density of 
$n_{0}=0.002h_{50}^{3}{\rm Mpc^{-3}}$ (eg. Efstathiou et al. 1988), 
these scale radii correspond to a range for $\tau_{g}$ (equation \ref{tg2}):
$0.01\simlt\tau_{g}\simlt0.18$. 
These ranges are consistent with those assumed 
in the intervening galaxy obscuration models of
Heisler \& Ostriker (1988) and Fall \& Pei (1993).

The values for ($\delta$, $z_{dust}$) were 
chosen to cover a range of evolution strengths for $r_{0}$ and
$\tau_{B}$ respectively. 
To cover a plausible range of dust formation epochs,
we consider $6\leq z_{dust}\leq20$,
consistent
with a range of galaxy `formation' epochs predicted by existing 
theories of structure formation (eg. Peebles 1989).
The upper bound $z_{dust}=20$ corresponds to 
the star formation epoch considered in the galaxy formation models 
of Blain \& Longair (1993b).

We assume values for $\delta$ similar to those implied 
by observations of the space density of
metal
absorption systems from QSO spectra as a
function of redshift (Sargent, Boksenberg \& Steidel 1988; Thomas \&
Webster 1990).
These systems are thought to arise in
gas associated with galaxies and their haloes and it is quite 
plausible that such systems also contain dust.
Here we assume a direct proportionality between the amount of dust and
heavy metal abundance in these systems. 

In general, evolution in the number of metal absorption
line systems per unit $z$, that takes into account
effects of cosmological expansion, can be parameterised:
\begin{equation}
\frac{dN}{dz}\,=\,\frac{c}{H_{0}}n_{z}\pi r_{0}(z)^{2}
(1+z)(1+2q_{0}z)^{-1/2}.
\label{linev}
\end{equation}
Evolution, such as a reduction in absorber numbers with redshift,
can be interpreted as either a decrease in the comoving
number density $n_{z}$, or effective cross-section $\pi r_{0}(z)^{2}$.
With our assumption of
a constant comoving density $n(z)=n_{0}$, and
an evolving dust scale radius $r_{0}$ as
defined by equation (\ref{rozev}), we have
$dN/dz\propto(1+z)^{\gamma}$, where
$\gamma=0.5+2\delta$ for $q_{0}=0.5$.
Hence for {\it no evolution}, $\gamma=0.5$.

Present estimates on the evolution
of absorber numbers with redshift are poorly constrained.
Thomas \& Webster (1990) have combined several datasets increasing
absorption redshift ranges to give
strong constraints on evolution models.
For C{\small IV} absorption ($\lambda\lambda$1548, 1551\AA), 
which can be detected to redshifts $z\simgt3$ in high resolution optical
spectra, evolution has been confirmed for the highest equivalent
width systems with $W_{0}\simgt0.6$\AA. 
It is more likely that these systems are those associated with dust rather than 
the lower equivalent width (presumably less chemically enriched) 
systems with $W_{0}\simlt0.3$\AA which have a trend
consistent
with no evolution.
Their value for the evolution parameter $\gamma$, for the
highest equivalent width systems is $-0.1\pm0.5$ at the $2\sigma$ level. 
Converting this $2\sigma$ range to our 
model parameter $\delta$ using 
the discussion above, we assume
the range: $-0.5<\delta<-0.05$. 

\subsection{Comparisons with QSO Absorption-Line Studies}

We can compare our assumed ranges in evolutionary parameters:
$6\leq z_{dust}\leq20$ and $-0.5<\delta<-0.05$
with recent determinations of the heavy element abundance
in damped Ly-$\alpha$ absorption
systems and the Ly-$\alpha$ forest to $z\sim3$.
The damped Ly-$\alpha$ systems are interpreted as the
progenitors of galactic disks (Wolfe et al. 1986), and recent
studies by Pettini et al. (1994; 1997)
deduce metal abundances and dust-to-gas ratios at $z\sim1.8-2.2$
that are
$\sim 10\%$ of the local value.
The Lyman forest systems however are more numerous, and
usually correspond to gas columns
$>10^{7}$ times lower than those of damped Ly-$\alpha$ absorbers.
High resolution metal-line observations by Songaila (1997)
deduce metallicities $\simlt1.5\%$ solar at $z\sim2.5-3.8$.

\begin{figure}
\vspace{-1in}
\plotonesmall{1.1}{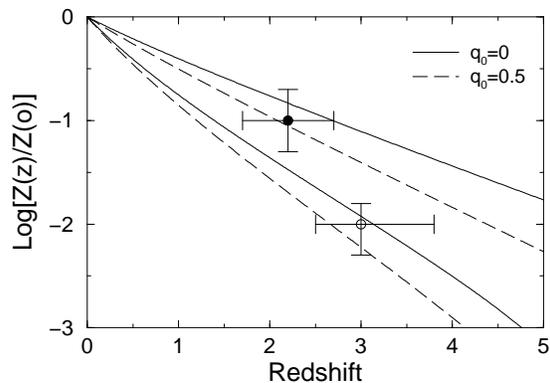}
\vspace{-1.5in}
\caption[]{Relative metallicity as a function of redshift. 
Regions within the solid and dashed curves represent 
the ranges predicted by our model for $q_{0}=0$ and $q_{0}=0.5$ respectively.
The filled and open data points
with $1\sigma$ error bars are mean observed values
from Pettini et al. (1994) and Songaila (1997) respectively.}
\label{metfit}
\end{figure}

To relate these metallicity estimates to cosmic evolution in dust
content as specified by our model, we must first note that 
the metallicity at any redshift $Z(z)$, is generally
defined as the mass fraction of heavy metals relative
to the total gas mass: $Z(z)=\Omega_{m}(z)/\Omega_{g}(z)$.
At all redshifts, we assume a constant dust-to-metals ratio,
$\Omega_{d}(z)/\Omega_{m}(z)$, where a fixed fraction of
heavy elements is assumed to be condensed into dust grains.
Therefore the metallicity $Z(z)$, relative to the local solar
value, $Z_{\odot}$, can be written:
\begin{equation}
\frac{Z(z)}{Z_{\odot}}\,=\,\frac{\Omega_{d}(z)}{\Omega_{d}(0)}
\frac{\Omega_{g}(0)}{\Omega_{g}(z)}.
\label{mdg}
\end{equation}
From the formalism in section~\ref{ev},
the mass density in dust
relative to the local density, $\Omega_{d}(z)/\Omega_{d}(0)$,
can be determined and is found to be
independent of the galaxy properties $r_{0}$ and $\tau_{B}$,
depending only on
our evolution parameters, $\delta$ and $z_{dust}$.
This is given by
\begin{equation}
\frac{\Omega_{d}(z)}{\Omega_{d}(0)}\,=\,\left[1 -
\frac{\ln(1+z)}{\ln(1+z_{dust})}\right](1+z)^{2\delta}.
\label{omegev}
\end{equation}
The gas ratio, $\Omega_{g}(0)/\Omega_{g}(z)$, is adopted
from studies of the evolution in gas content of damped Ly-$\alpha$ systems.
These systems are believed to account for at least $80\%$
of the gas content in the form of
neutral hydrogen at redshifts $z\simgt2$ (Lanzetta et al. 1991).
We adopt the empirical fit of Lanzetta et al. (1995), who find that the
observed evolution in $\Omega_{g}(z)$ is
well represented by $\Omega_{g}(z)=\Omega_{g}(0)\exp(\alpha z)$,
where $\alpha=0.6\pm0.15$ and $0.83\pm0.15$ for
$q_{0}=0$ and $q_{0}=0.5$ respectively.

\begin{figure*}
\vspace{-0.6in}
\plotonesmall{1.4}{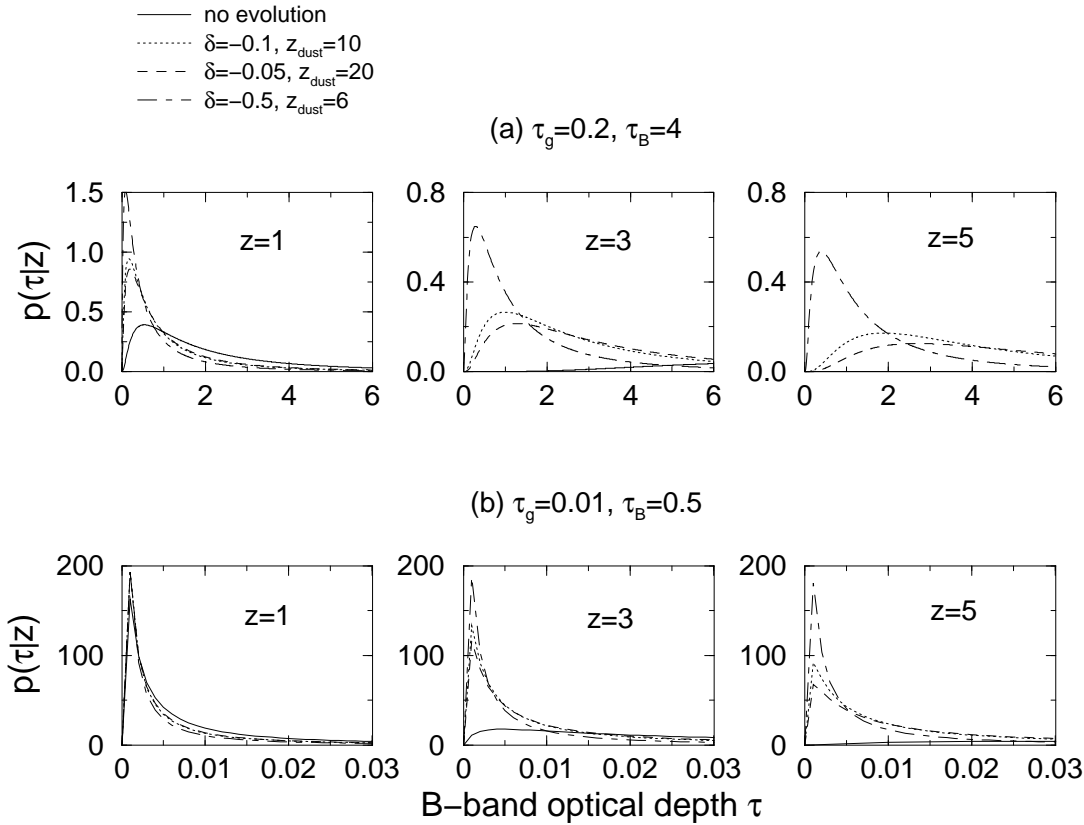}
\vspace{-1in}
\caption[Optical depth probability distribution functions in evolutionary model]
{Optical
depth probability distribution functions $p(\tau|z)$
to redshifts $z=$1, 3 and 5, where $\tau$ is the total optical depth
observed in the $B$-band.
Two different sets of galaxy parameters ($\tau_{g}$, $\tau_{B}$) are considered:
(a) (0.2,4) and (b) (0.01,0.5) (see section~\ref{param}).
For each of these, we show four evolutionary models specified by
($\delta$, $z_{dust}$). `No evolution' corresponds to $\delta=0$ and
$z_{dust}=\infty$ and the `Strongest evolution' to
$\delta=-0.5$ and $z_{dust}=6$.}
\label{pdfs2}
\end{figure*}

Figure~\ref{metfit} shows the range in relative metallicity implied by our
evolutionary dust model (equations~\ref{mdg} and~\ref{omegev}) as
a function of redshift for two values of $q_{0}$. 
The solid and dashed lines correspond to respectively $q_{0}=0$ and
$q_{0}=0.5$ and the regions between these lines
correspond to the ranges
assumed for our assumed model parameters: $6\leq z_{dust}\leq20$ and 
$-0.5<\delta<-0.05$.
For comparison, the mean metallicities
$Z\approx 0.1Z_{\odot}$ and $Z\approx 0.01Z_{\odot}$
observed in damped Ly-$\alpha$ systems at
$z\approx2.2$ and the Lyman forest at $z\simgt2.5$ respectively
are also shown. These agree well with our model predictions,
suggesting that our model assumptions will provide a reliable measure
of dust evolution which are at least compatible with other indirect estimates. 

\section{Results and Analysis}
\label{rests}

Using the formalism of Masci \& Webster (1995) and replacing
the parameters $\tau_{B}$ and $r_{0}$ by their assumed redshift dependence
as defined in section~\ref{ev},
Fig.~\ref{pdfs2} shows
probability density functions $p(\tau\, |\,z)$ for the
total optical depth up to redshifts $z=$1, 3 and 5. Results are shown 
for two sets of galaxy parameters ($\tau_{g}$, $\tau_{B}$), with four sets of
evolutionary parameters ($\delta$, $z_{dust}$) for each.
 
The area under any normalised 
curve in Fig.~\ref{pdfs2} gives the fraction
of lines-of-sight to that redshift which have optical
depths within some interval $0\rightarrow\tau_{max}$. 
Towards high
redshifts, we find that 
obscuration depends most sensitively on the parameter $\tau_{g}$, in
other words, on the covering factor of absorbers
(equation \ref{tg2}). 
Figure~\ref{pdfs2} shows that as the amount
of dust at high redshift decreases, ie., as $\delta$ and $z_{dust}$ 
decrease,
the curves show little horizontal shift
towards larger optical depths from $z=1$ to $z=5$.
A significant shift becomes 
noticeable however for the weaker evolution cases, and is largest 
for `no evolution' (solid lines). 
This behaviour is further investigated below. 

In order to give a clearer comparison between the 
amount of obscuration and strength of evolution implied by our model
parameters ($\tau_{g},\,\tau_{B},\,\delta,\,z_{dust}$), 
we have calculated the mean and variance in
total optical depth as a function of redshift.
Formal derivations of these quantities are
given in 
the appendix. Here we briefly discuss their general dependence
on the model parameters. 

\begin{figure*}
\vspace{-0.5in}
\plotonesmall{1.4}{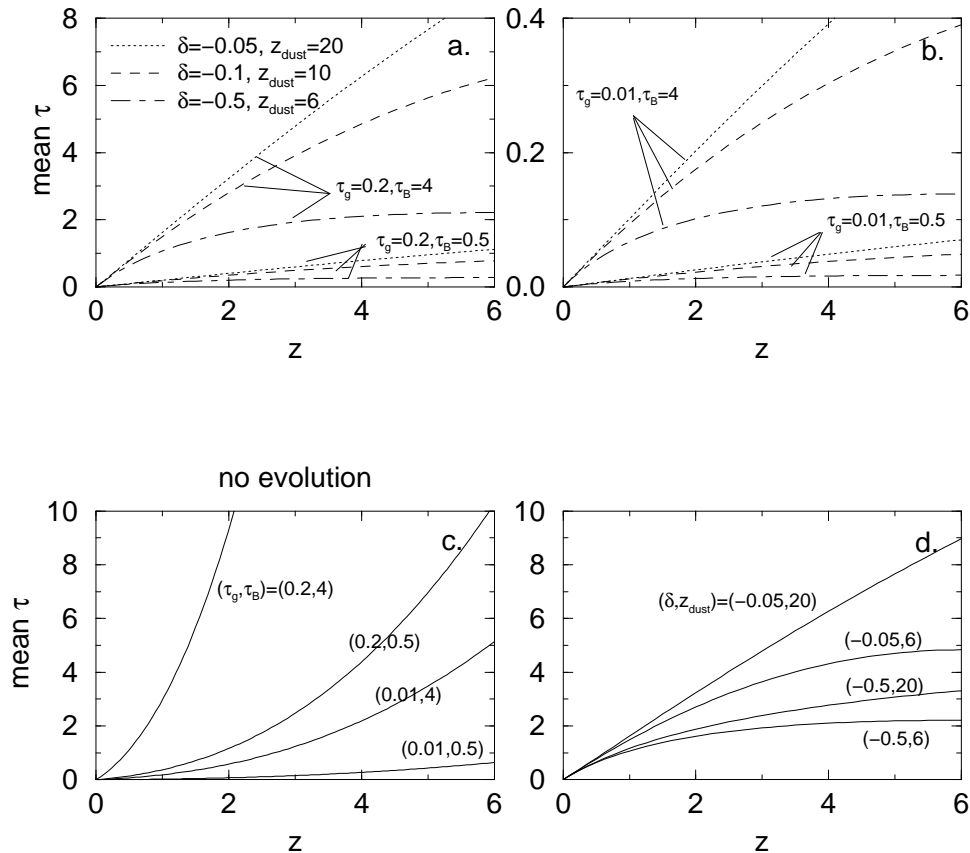}
\vspace{-1in}
\caption[Mean optical depth as a function of $z$ in evolutionary model]{
Behaviour in mean reddening,
$\langle\tau\rangle$, as a function of
redshift for a range of model parameters ($\tau_{g}$, $\tau_{B}$)
and ($\delta$, $z_{dust}$).
(a) For ($\tau_{g},\tau_{B}$)=(0.2,4) and (0.2,0.5), (b) Same as (a) but for
$\tau_{g}=0.01$,
(c) Redshift dependence of mean reddening in {\it no-evolution} model
for a range of parameters ($\tau_{g}$, $\tau_{B}$).
(d) Scaling of the mean reddening with respect to the evolutionary
parameters ($\delta$, $z_{dust}$) with ($\tau_{g}$, $\tau_{B}$)
fixed at (0.2,4).}
\label{meantaus}
\end{figure*}

A quantity first worth considering is the number of galaxies
intercepted along the line-of-sight. In a $q_{0}=0.5$ ($\Lambda=0$)
universe, the average number of intersections
within a scale length $r_{0}$ of a galaxy's center by a light ray
to some redshift is given by 
\begin{equation}
\bar{N}(z)\,=\,\left({2\over3+4\delta}\right)\tau_{g}
\left[(1+z)^{1.5\,+\,2\delta}
- 1\right].
\label{nzev}
\end{equation}
Where $\delta$ and $\tau_{g}$ are defined in equations (\ref{rozev})
and (\ref{tg2}) 
respectively. 
 
In the case where we have {\it no-evolution}, ie. where $\delta=0$ and $z_{max}=\infty$, and for a dust law that scales inversely with wavelength
(ie. $\xi_{\lambda}\propto 1/\lambda$ which is a good approximation at
$\lambda\simgt2500$\AA), 
exact expressions follow for the mean and variance in total optical depth
along the line-of-sight.
The mean optical depth can be written: 
\begin{equation}
\bar{\tau}(z)\,=\,0.8\,\tau_{g}\,\tau_{B}\left[(1+z)^{2.5} -
1\right],
\label{taunoev}
\end{equation}
and the variance:
\begin{equation}
\sigma^{2}_{\tau}(z)\,=\,
0.57\,\tau_{g}\,{\tau_{B}}^{2}\,\left[(1+z)^{3.5} - 1\right].
\label{varnoev}
\end{equation}
 
The variance (equation \ref{varnoev}) or `scatter' about the mean to some redshift
provides a more convenient measure of reddening. 
The mean optical depth has a simple linear dependence on the parameters
$\tau_{g}$ and $\tau_{B}$ and thus gives no indication of the degree
to which each of these parameters contributes to the scatter. 
As seen from the probability distributions
in Fig.~\ref{pdfs2}, 
there is a relatively large scatter about the mean optical depth 
to any
redshift. 
From equation (\ref{varnoev}), 
it is seen that the strongest dependence 
of the variance is on the central absorber optical depth $\tau_{B}$.
Thus, larger values of $\tau_{B}$ (which imply `harder-edged' disks), 
are expected
to introduce considerable scatter amongst random individual
lines of sight, even to relatively low redshift.  

In Fig.~\ref{meantaus}, we show how the mean optical depth varies as a
function of redshift for a range of evolutionary parameters.
`Strong evolution' is characterised by $\delta=-0.5$, $z_{dust}=6$
(dot-dashed curves),
as compared to the `no', `weak' and
`moderate' evolution cases indicated.
The mean optical depth
flattens out considerably towards high redshift in the strong evolution case, 
and 
gradually steepens
as $\delta$ and $z_{dust}$ are increased.
Note that no such flattening is expected in mean reddening for the
no evolution case (Fig.~\ref{meantaus}c).
The mean optical depth to redshifts $z\simgt1$ in evolution models can be 
reduced by factors of at least three, even for low to moderately low
evolution strengths. 
 
\begin{figure}
\vspace{-0.8in}
\plotonesmall{1.1}{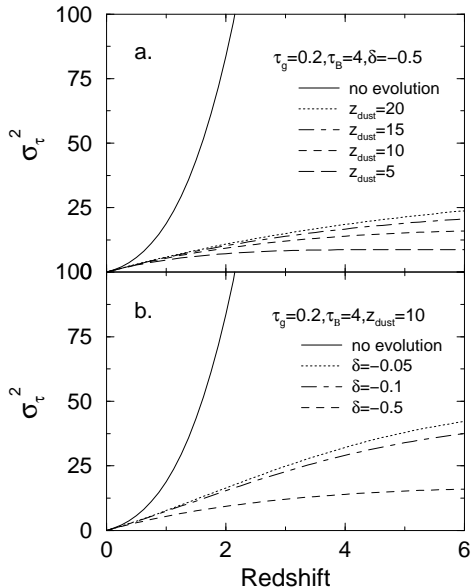}
\vspace{-0.8in}
\caption[Variance in optical depth as a function of redshift]{
Variance ($\sigma_{\tau}^{2}$) in optical depth
as a function of redshift showing scaling with respect to
the evolution parameters ($\delta$, $z_{dust}$). ($\tau_{g},\tau_{B}$)
are fixed at (0.2,4). (a) $\delta$ fixed at -0.05 and
$z_{dust}$ is varied.
(b) $z_{dust}$ fixed at 10 and $\delta$ is varied.
}
\label{vars}
\end{figure}

Figure~\ref{meantaus}d 
shows the scaling of mean optical depth with respect to
the evolutionary parameters. 
It is seen that reddening depends most sensitively on the parameter
$\delta$, which controls the rate of evolution in galactic dust scale radius 
$r_{0}$.
A similar trend is followed in Fig.~\ref{vars}, which shows the dependence of
variance in optical depth on evolution as a function of redshift,
for fixed ($\tau_{g}$, $\tau_{B}$). Considerable
scatter is expected if the dust radius of a typical galaxy evolves
slowly with cosmic time as shown for the `weakest' 
evolution case $\delta=-0.05$ in Fig.~\ref{vars}. 
 
Our main conclusion is that the inclusion of evolution in dust content,
by amounts
consistent with other indirect studies can dramatically reduce
the redshift dependence of total reddening along the line-of-sight
to $z\simgt1$, contrary 
to non-evolving models. 

\section{Implications on QSO Number Counts}
\label{QSOev}
 
There are numerous observations suggesting that the
space density
of bright quasars declines beyond $z\approx3$
(Sandage 1972; Schmidt, Schneider \& Gunn 1988).
This has been strongly confirmed from various 
luminosity function (LF) estimates to $z\sim4.5$ (Hartwick \& Schade 1990;
Pei 1995 and references therein), where 
the space density is seen to decline by at least an order of magnitude
from $z=3$ to $z=4$.
Heisler \& Ostriker (1988) speculate that the decline may be due to
obscuration by intervening dust, which reduces the number of quasars
observed by ever-increasing amounts towards high $z$. 
The results of Fall \& Pei (1993) however show that the observed
turnover at $z\sim2.5$ and decline thereafter may still exist once
the effects of intervening dust (mainly associated with damped Ly$\alpha$
systems) are corrected for. 
Since no evolution in dust content was assumed in either of these studies,
we shall further explore the effects of intervening dust on 
inferred quasar evolution using our evolutionary galactic dust model.
 
Since we are mainly interested in ``bright'' quasars ($M_{B}\simlt-26$)
at high redshifts, a single power-law for the observed LF should suffice:
\begin{equation}
\label{oLF}
\phi_{o}(L,z)\,=\,\phi_{\ast o}(z)L^{-\beta-1},
\end{equation} 
where $\beta\simeq2.5$.
This power law model immensely simplifies the relation 
between observed and ``true'' LFs (corrected for obscuration by dust).
In the presence of dust obscuration, inferred luminosities will be
decreased by a factor of $e^{-\tau}$. Since there is a probability $p(\tau\,|\,z)$ 
of encountering an optical depth $\tau$ as specified
by our model (see Fig.~\ref{pdfs2}),
the observed LF can be written in
terms of the true LF, $\phi_{t}$ as follows:
\begin{equation}
\label{oLFtLF}
\phi_{o}(L,z)\,=\,\int^{\infty}_{0}d\tau\,\phi_{t}(e^{\tau}L,z)
e^{\tau}p(\tau\,|\,z)
\end{equation}
The extra factor of $e^{\tau}$ in equation (\ref{oLFtLF})
accounts for a decrease in luminosity 
interval $dL$ in the presence of dust. 
Equations (\ref{oLF}) and (\ref{oLFtLF}) imply that the true LF can be written 
\begin{equation}
\label{tLF}
\phi_{t}(L,z)\,=\,\phi_{\ast t}(z)L^{-\beta-1},
\end{equation}
and the ratio of observed to true LF normalisation as
\begin{equation} 
\label{norm}
\frac{\phi_{\ast o}(z)}{\phi_{\ast t}(z)}\,=\,\int^{\infty}_{0}d\tau\,
e^{-\beta\tau}p(\tau\,|\,z). 
\end{equation}

The observed comoving density of quasars brighter than some
absolute magnitude limit $M_{lim}$ as a function of redshift is
computed by integrating the LF:
\begin{equation}
\label{No}
N_{o}(z\,|M_{B}<M_{lim})\,=\,\int^{\infty}_{L_{lim=L(M_{lim})}}dL\,\phi_{o}(L,z).
\end{equation} 
Thus, the true comoving number density $N_{t}$, can be easily
calculated by replacing $\phi_{o}$ in equation (\ref{No}) by
$\phi_{t}\equiv(\phi_{\ast t}/\phi_{\ast o})\phi_{o}$ leading to the
simple result:
\begin{equation}
\label{Nt}
N_{t}(z\,|M_{B}<M_{lim})\,\simeq\,\left(\frac{\phi_{\ast o}(z)}{\phi_{\ast t}(z)}
\right)N_{o}(z\,|M_{B}<M_{lim}),
\end{equation} 
where the normalisation ratio is defined by equation (\ref{norm}). 
 
\begin{figure}
\vspace{-0.5in}
\plotonesmall{1.1}{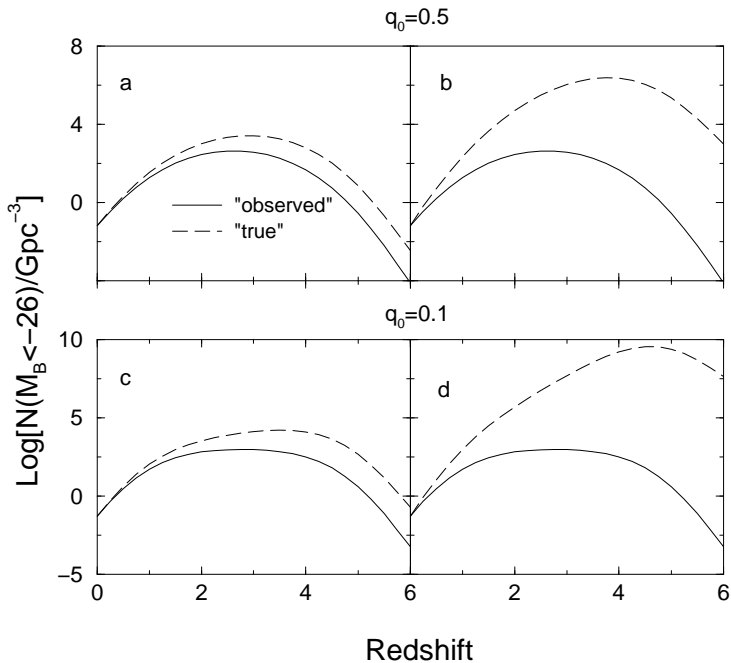}
\vspace{-0.3in}
\caption[Comoving number density of quasars with $M_{B}<-26$ as a function of redshift]{ 
Comoving number density of quasars with $M_{B}<-26$ 
as a function of redshift. Observed trends (solid curves)
are taken from the empirical fits of 
Pei (1995) while dashed curves corrects these trends 
for obscuration by dust. These are predicted assuming 
our
evolving intervening
galactic dust model with $\tau_{B}=1$ and $r_{0}=10$kpc (Figs a
and c) and $\tau_{B}=3$ and $r_{0}=30$kpc (Figs b and d). 
In all cases, we have assumed relatively ``weak'' evolution in 
dust content with $z$, defined by the parameters: 
$z_{dust}=20$ and $\delta=-0.05$. 
}
\label{comov}
\end{figure}

Figure~\ref{comov} shows both the observed and true comoving 
density of bright quasars (with $M_{B}<-26$) as a function of redshift. 
The observed trends are empirical fits deduced by Pei (1995). 
The true comoving density in all cases was determined by assuming relatively
`weak' evolution in the dust properties of intervening galaxies. 
Two sets of galactic dust parameters for each $q_{0}$ defined by
$(\tau_{B},r_{0})=(1,10{\rm kpc})$ (Figs a and c) and 
$(\tau_{B},r_{0})=(3,30{\rm kpc})$
(Figs b and d) are assumed. We shall refer to these as our 
``minimal'' and ``maximal'' dust model respectively which bracket the
range of parameters observed for local galaxies. 
 
Comparing the `true' QSO redshift distribution with that observed,
two features are apparent.
First, the true number density vs. $z$ relation has qualitatively
the same behaviour as that observed.
No flattening or increase in true quasar numbers with $z$ is apparent. 
Second, there appears to be a shift in the redshift, $z_{peak}$,
where the quasar density peaks. This shift is
greatest for our maximal dust model where $z_{peak}$ is increased
by a factor of almost 1.5 relative to that observed. 
This implies that the bulk of quasars may have formed at earlier
epochs than previously inferred from direct observation.
 
Our predictions for QSO evolution, corrected for obscuration
by `evolving' intervening dust differs
enormously from that predicted by Heisler \& Ostriker (1988).
The major difference is that these authors neglected evolution in dust
content with $z$. As shown in Fig.~\ref{meantaus}, 
non-evolving models lead to a rapid increase
in dust optical depth with $z$ and hence this will explain 
their claim of a continuous increase in the true QSO space density
at $z>3$.
As shown in Fig.~\ref{comov}, the inclusion of even a low-to-moderately low 
amount of evolution in dust content dramatically reduces the excess number
of quasars at $z>3$ than predicted by Heisler \& Ostriker (1988). 
 
We find that there is no significant difference in the characteristic
timescale, $t_{QSO}$ for QSO formation at $z>z_{peak}$, where
\begin{equation}
\label{time}
t_{QSO}\,\simeq\,\left(\frac{N}{\stackrel{.}N}\right)_{z>z_{peak}}\,\sim\,
1.5\,{\rm Gyr}, 
\end{equation} 
is found for both the observed and dust corrected results in Fig.~\ref{comov}.
We conclude that the decline in space density of bright QSOs
at redshifts $z>3.5$ is most likely to be real and an artifact
of an intrinsic rapid turn-on of the QSO population with time. 
This is consistent with estimates of evolution inferred from
radio-quasar surveys where no bias from 
dust obscuration is expected 
(eg. Dunlop \& Peacock 1990). 
 
An increased space
density of quasars 
at redshifts $z>3$ predicted by correcting for dust obscuration
has implications for theories of structure formation in the Universe.
Our minimal dust model 
(Figs.~\ref{comov}a and c) predicts that the true space density
can be greater by almost two orders of magnitude than that observed, while our
maximal dust model (Figs.~\ref{comov}b and d) predicts this
factor to be greater than 5 orders of magnitude. 
These predictions can be reconciled with the quasar number densities
predicted from hierarchical galaxy formation simulations involving 
cold-dark matter
(eg. Katz et al. 1994). 
It is found that there are $>10^{3}$ times
potential quasar sites at $z>4$ (associated with high density peaks) 
than required from current observations.
Such numbers can be 
easily accommodated by our predictions if a 
significant quantity of line-of-sight dust is present. 
 
To summarise, we have shown that with the inclusion of even weak
to moderately weak
amounts of evolution in dust content with $z$, the bias due
to dust obscuration will not be enough to flatten the true 
redshift distribution of bright quasars beyond $z=3$. 
A significant excess however (over that observed) in quasar numbers is 
still predicted.

\section{Discussion}
\label{evmodd}
 
Our model predictions may critically depend 
on the dust properties
of individual galaxies and their assumed evolution.
For instance, is it reasonable to give galaxies an exponential
dust distribution? Such a distribution is expected to give
a dust covering factor to some redshift considerably larger
than if a clumpy distribution
were assumed (Wright, 1986).
A clumpy dust distribution (for spirals
in particular) is expected, as dust is
known to primarily form in dense, molecular star-forming
clouds (Wang 1991 and references therein).

As noted by Wright (1986), ``cloudy disks'' 
with dust in optically-thick clumps
can reduce the
effective cross-section for dust absorption by at least 
a factor of five and hence, are
less efficient at both obscuring and reddening background sources
at high redshift.
A dependence of the degree of dust `clumpiness'
on redshift, such as dust which is more diffuse at early epochs 
and becomes more clumpy with cosmic time
is unlikely to affect the results of this paper. 
This will only reduce the effective cross-section for absorption 
to low redshifts, leaving the effects to high redshift essentially 
unchanged.
The numbers of reddened and/or obscured sources at high redshift
relative to those expected in non-evolving dust models however
will always be reduced, regardless of the dependence of 
aborption cross-section on redshift.

Observations of the optical reddening distribution
of quasars as a function of 
redshift may be used to test our predictions. Large and complete
radio-selected samples with a high identification rate 
extending to high redshifts however
are required. The reason for this is that first,
radio wavelegths are guaranteed to have no bias against obscuration by dust,
and second, the statistics at high redshift need to be reasonably
high in order to provide sufficient sampling 
of an unbiased number of random sight-lines. 

The sample of Drinkwater et al. (1997) contains
the highest quasar fraction ($\simgt70\%$) than any existing radio sample 
with a redshift distribution extending to $z\sim4$.
A large fraction of sources appear very red in $B-K$ colour compared 
to quasars selected by optical means.
The dependence of $B-K$ colour on redshift 
is relatively flat which may at first appear
consistent with the predictions of figure~\ref{meantaus}, 
although the fraction of sources identified 
with $z\simgt2$ is only $\sim5\%$.
Also, this sample is known to contain large
numbers of sources which are reddened by mechanisms other than dust
in the line-of-sight (eg. Serjent \& Rawlings 1996).
The role of dust, reddening the optical--to--near-IR continua of radio-selected
quasars, and whether it is extrinsic or not still
remains a controversial issue.
One needs to isolate the intrinsic source properties before attributing
any excess reddening to line-of-sight dust. 
Optical follow-up of sensitive radio surveys that detect
large numbers of high redshift sources with known intrinsic spectral
properties will be necessary to reliably constrain the
rate of evolution in cosmic dust.
 
\section{Summary and Conclusions}
\label{concfour}
 
In this paper, we have modelled the optical depth in galactic dust 
along the line-of-sight as a function of redshift
assuming evolution in dust content.
Our model depends on four parameters which specify the 
dust properties of local galaxies and their evolution:
the exponential dust scale radius $r_{0}$, central $B$-band optical
depth $\tau_{B}$, ``evolution strength'' $\delta$ where
$r_{0}(z)=r_{0}(1+z)^{\delta}$, and $z_{dust}$ - a hypothesised
dust formation epoch.
Our evolution model is based on previous studies of the formation
of heavy metals in the cold dark matter scenario of galaxy formation. 
 
Our main results are:
\\\indent 1. For evolutionary parameters 
consistent with existing studies of the evolution of metallicity
deduced from
QSO absorption-line systems, 
a significant ``flattening'' in the mean and variance
of observed $B$-band optical depth
to redshifts $z>1$ is expected.
The mean optical depth to $z\simgt1$
is smaller by at least a factor of 3 compared to
non-evolving model predictions. Obscuration by dust is not
as severe as shown in previous studies if effects of evolution are accounted 
for.
\\\indent 2. By allowing for even moderately low amounts of evolution,
line-of-sight dust is {\it not} expected 
to significantly affect existing optical studies of QSO evolution.
Correcting for dust obscuration, evolving dust models
predict the `true' (intrinsic) space density of bright quasars to
decrease beyond $z\sim2.5$ as observed, contrary to 
previous non-evolving dust models where a continuous monotonic
increase was predicted.
\\\indent 3. For moderate amounts of evolution, our models 
predict a mean observed $B$-band optical depth that scales
as a function of redshift as $\bar{\tau}\propto(1+z)^{0.1}$. 
For comparison, evolving models predict a dependence:
$\bar{\tau}\propto(1+z)^{2.5}$. We believe
future radio surveys of high
sensitivity that reveal large numbers of optically reddened sources
at high redshift will provide the necessary data to constrain 
these models.

\section{Acknowledgments}
The authors would like to thank Paul Francis for many 
illuminating discussions and the referee for providing valuable
suggestions on the structure of this paper.
FJM acknowledges support from an Australian Postgraduate Award.

\appendix
\section{Derivation of Mean Optical Depth}

Here we derive expressions for the mean and variance in 
total optical depth as a function of redshift in
our evolutionary galactic dust model discussed in section~\ref{resan}.
The galaxies are modelled as exponential dusty disks, randomly
inclined to the line-of-sight. 

We first derive the average number of galaxies intercepted by a light ray
emitted from some redshift $z$ (ie. equation \ref{nzev}).
Given a `proper' number density of galaxies at some redshift
$n_{g}(z)$, with each galaxy having an effective cross-sectional area
$\mu\,\sigma$ as viewed by an observer
($\mu$ is a random inclination factor, where
$\mu=\cos\,\theta$ and $\theta$ is the angle
between the sky plane and the plane of a galactic disk),
the average number of intersections of a light ray along some path
length $ds$ will be given by 
\begin{equation}
dN\,=\,n_{g}(z)\,\mu\sigma\,ds.
\label{dn}
\end{equation}
In an expanding universe we have $n_{g}=n_{0}(1+z)^3$, where $n_{0}$ is
a local comoving number density and is assumed to be constant. Units of proper
length
and redshift are related by
\begin{equation}
{ds\over dz}\,=\,\left({c\over
H_{0}}\right)\,{1\over(1+z)^{2}(1+2q_{0}z)^{1/2}}
\label{ds}
\end{equation}
(Weinberg 1972). The effective cross-section projected towards an
observer
for a randomly inclined disk is found by averaging over the random
inclination factor
$\mu$, where $\mu$ is randomly distributed between 0 and 1, and
integrating
over the exponential profile assumed for each disk with scale radius
$r_{0}(z)$ (see equations \ref{expr} and \ref{rozev}).
The product $\mu\sigma$ in equation (\ref{dn}) is thus replaced by
\begin{equation}
\int_{0}^{1}\mu\,d\mu\,\int_{0}^{\infty}e^{-r/r_{0}(z)}\,
2\pi r\,dr\,=\,\pi\,{r_{0}}^2(1+z)^{2\delta}.
\label{eq}
\end{equation}
Thus from equation (\ref{dn}), the average number of intersections to some
redshift $z$ is given by
\begin{eqnarray}
\nonumber
\bar{N}(z)\,& = &\,\int_{0}^{z}\mu\sigma\,n_{g}(z')\,\left({ds\over
dz'}\right)\,dz'\\ 
& = &\,n_{0}\pi r_{0}^{2}\left({c\over
H_{0}}\right)\int_{0}^{z}{(1+z')^{1+2\delta
}\over (1+2q_{0}z')^{1/2}}\,dz'.
\label{n}
\end{eqnarray}
With $\tau_{g}$ defined by $n_{0}\pi r_{0}^{2}\left({c\over
H_{0}}\right)$,
this directly leads to equation (\ref{nzev}) for $q_{0}=0.5$.
 
The mean optical depth $\bar{\tau}$ is derived by a similar argument.
If $\tau_{0}(z)$ is the optical depth observed through a face on galaxy
at
some redshift $z$ (equation \ref{tboz}), then a galactic disk inclined by some
factor 
$\mu$ will
have its optical depth increased to $\tau_{0}(z)/\mu$. Multiplying this
quantity
by equation (\ref{dn}), the extinction suffered by a light ray along a path
length $ds$
is given by
\begin{equation}
d\tau\,=\,n_{g}(z)\,\sigma\,\tau_{0}(z)\,ds.
\label{dt}
\end{equation}
Thus the mean optical depth to some redshift $z$ can be calculated from
\begin{equation}
\bar{\tau}(z)\,=\,\int_{0}^{z}\sigma n_{g}(z')\tau_{0}(z')\left({ds\over
dz'}\right
)\,dz'.
\label{mt}
\end{equation}
Given $n_{g}(z)$, $\left({ds\over dz}\right)$ and $\sigma$ (from the
integral over $r$ in equation \ref{eq})
above,
and $\tau_{0}(z')$ from equation (\ref{tboz}), the mean optical depth follows the
general form
\begin{eqnarray}
\nonumber
\bar{\tau}(z)\,&=&\,2\tau_{g}\tau_{B}\int_{0}^{z}{(1+z')^{1+2\delta
}\over (1+2q_{0}z')^{1/2}}\,\xi\left(\frac{\lambda_{B}}{1+z}\right)\\
& \times &\left[1 -
{\ln(1+z')\over\ln(1+z_{dust})}\right]\,dz'.
\label{mtz}
\end{eqnarray}
Similarly, the variance in the optical depth distribution is defined as 
follows:
\begin{equation}
\sigma_{\tau}^{2}(z)\,=\,\langle\tau^2\rangle -
{\langle\tau\rangle}^{2}\,=\,\int_{0}^{z}\sigma
n_{g}(z'){\tau_{0}}^{2}(z')\left({ds\over dz'}\right)\,dz'.
\label{vart}
\end{equation}
\noindent
In terms of our model dependent parameters, this becomes
\begin{eqnarray}
\nonumber
\sigma_{\tau}^{2}(z)\,&=&\,2\tau_{g}{\tau_{B}}^{2}\int_{0}^{z}{(1+z')^{1+2\delta
}\over (1+2q_{0}z')^{1/2}}\,\xi^{2}\left(\frac{\lambda_{B}}{1+z}\right)\\
& \times& {\left[1
-{\ln(1+z')\over\ln(1+z_{dust})}\right]}^{2}\,dz'.
\label{vartz}
\end{eqnarray}

\end{document}